%% file: main.tex
\documentclass[conference,9pt]{IEEEtran}
\IEEEoverridecommandlockouts
\usepackage{cite}
\usepackage{amsmath,amssymb,amsfonts}
\usepackage{algorithmic}
\usepackage{graphicx}
\usepackage{textcomp}
\usepackage{xcolor}
\usepackage{syntax}
\usepackage{tikz}

\usepackage{multirow}
\usepackage{booktabs}

\usetikzlibrary{arrows.meta,positioning,fit,backgrounds,shapes.misc,calc}
\usepackage{listings}
\lstdefinelanguage{Lspec}{
  sensitive=true, morecomment=[l]{//}, morestring=[b]",
}
\lstdefinestyle{promptstyle}{
  basicstyle=\scriptsize\ttfamily, columns=fullflexible, keepspaces=true,
  breaklines=true, breakatwhitespace=true, breakindent=0pt, literate={-}{{-}}1,
  frame=single, backgroundcolor=\color{cyan!8},
  xleftmargin=3.4pt, xrightmargin=3.4pt, aboveskip=4pt, belowskip=4pt,
}
\lstdefinestyle{specstyle}{
  language=Lspec, basicstyle=\scriptsize\ttfamily, commentstyle=\color{gray}\itshape,
  columns=fullflexible, keepspaces=true, showstringspaces=false,
  mathescape=true, escapeinside={(*@}{@*)}, breaklines=false,
  frame=single, backgroundcolor=\color{violet!8},
  xleftmargin=3.4pt, xrightmargin=3.4pt, aboveskip=4pt, belowskip=4pt,
}
\lstdefinestyle{codestyle}{
  language=Lspec, basicstyle=\scriptsize\ttfamily, commentstyle=\color{gray}\itshape,
  columns=fullflexible, keepspaces=true, showstringspaces=false,
  mathescape=true, escapeinside={(*@}{@*)}, breaklines=false,
  frame=single, backgroundcolor=\color{gray!12},
  xleftmargin=3.4pt, xrightmargin=3.4pt, aboveskip=4pt, belowskip=4pt,
}

\newcommand{\Lspec}{\ensuremath{\mathcal{L}_{\mathit{spec}}}}
\newcommand{\Lcode}{\ensuremath{\mathcal{L}_{\mathit{code}}}}

\def\BibTeX{{\rm B\kern-.05em{\sc i\kern-.025em b}\kern-.08em
    T\kern-.1667em\lower.7ex\hbox{E}\kern-.125emX}}
\begin{document}

\title{Formally-Verifiable and Human-Interpretable RTL Generation with LLM-Guided Stepwise Refinement}

\author{\IEEEauthorblockN{You Li}
\IEEEauthorblockA{
\textit{The University of Texas at Austin}\\
you.li@msn.com}
\and
\IEEEauthorblockN{Samuel Mandell}
\IEEEauthorblockA{
\textit{The University of Texas at Austin}\\
shm923@my.utexas.edu}
\and
\IEEEauthorblockN{David Z. Pan}
\IEEEauthorblockA{
\textit{The University of Texas at Austin}\\
dpan@utexas.edu}
}

\maketitle

\begin{abstract}
Large language models (LLMs) are accelerating register-transfer level (RTL) development, but ensuring the correctness of generated designs remains a major challenge. Existing LLM-driven workflows rely heavily on simulation-based verification, which provides limited correctness guarantees, and often offer little insight into the design decisions behind the generated RTL. We present a framework for LLM-guided, correct-by-construction RTL generation that combines the generative capabilities of LLMs with the mathematical rigor of stepwise refinement. The framework translates natural-language requirements into a formal hardware contract and progressively refines the contract into an implementation. At each refinement step, the LLM proposes a design decision, while the refinement engine applies the corresponding transformation only after formally verifying its correctness. The resulting program is formally verified against the contract and then translated into synthesizable RTL. The recorded derivation allows every line of generated RTL to be traced back to the design decisions and requirements that motivated it. Beyond generating verified implementations, the framework provides reliable verification feedback and reusable design experience, laying the foundation for self-improving hardware generation.
\end{abstract}

\input{src/intro}
\input{src/background}
\input{src/specification}
\input{src/refinement}
\input{src/implementation}

\input{src/evaluation}
\input{src/conclusion}

\newpage
\bibliographystyle{IEEEtran}
\bibliography{ref}

\end{document}

%% file: src/intro.tex
\section{Introduction}
\label{sec:intro}

Large language models (LLMs) are changing the way register-transfer level (RTL) designs are developed. Given a natural-language specification, an LLM agent can draft, simulate, and debug a module within minutes~\cite{zhao2025mage,liu2026veritrace}. As these agents become faster and more capable, the central challenge shifts from generating designs to establishing their correctness. A single subtle error can cause a silicon respin, incurring millions of dollars in mask costs at advanced technology nodes and delaying time to market. The primary safeguard in current LLM-driven design flows is simulation-based verification. However, it offers only limited assurance: 44.2\% of LLM-generated designs that pass all testbenches provided in the RTLLM benchmark suite fail formal verification~\cite{jin2025realbench}.

Hardware design and verification pose unique challenges for LLM agents. First, there is a fundamental mismatch between the sequential generation process of LLMs and the concurrent semantics of hardware. LLMs generate code token by token, whereas RTL describes concurrent processes with interdependent signals and register updates synchronized by clock edges. Generation accuracy drops sharply on problems involving control logic~\cite{liu2026howllmsfail}, finite-state machines~\cite{wu2026llmfsm}, and behavior spanning multiple clock cycles~\cite{zhang2026rtlerrors}. Second, LLMs tend to generate complete designs in a single pass, leaving the underlying design decisions implicit. Recent studies have introduced intermediate reasoning steps, but these steps either lack formal semantics or provide no correctness guarantees (Section~\ref{sec:background}). Without explicit and verifiable design decisions, reviewers must reconstruct design intent from the implementation, while verification engineers must manually trace errors without a record.

Program refinement provides a systematic framework that makes each design decision explicit and each design step verifiable~\cite{wirth1971program}. It progressively transforms a specification into an implementation through a sequence of steps, each governed by a refinement law. The resulting implementation is therefore correct by construction, \textit{i.e.}, it is guaranteed to satisfy the original specification. Despite its promise, program refinement has seen limited adoption in hardware design because it requires substantial human effort to guide the derivation. We address this limitation by using an LLM to make design choices and a refinement engine to apply the corresponding transformations and check their correctness.

In this paper, we propose an automated RTL development framework that combines the generative capabilities of LLMs with the mathematical rigor of program refinement. The framework consists of three stages. First, it translates the natural-language requirements into a formal hardware contract expressed in Dafny~\cite{leino2010dafny}, using deterministic checks to establish its formal validity and a semantic judge to assess its agreement with the original requirements.
Second, it progressively refines the contract into concrete code by applying refinement laws. To make a design decision, the LLM fills in the placeholders in the template of a refinement law. The refinement engine then applies the corresponding transformation and invokes Dafny to check the proof obligations specified by the template. If the derivation reaches a dead end, the engine backtracks and explores another choice.
Third, the concrete program is compiled to synthesizable RTL.
By separating design decisions from program transformations and verifying each refinement step, the framework makes the design process interpretable and ensures that the resulting program satisfies its formal contract.
To the best of our knowledge, this is the first framework for LLM-guided, correct-by-construction RTL generation.

Beyond generating individual hardware designs, our framework can pave the way toward recursive self-improvement (RSI) by providing reliable feedback and reusable design experience. Self-improvement requires a system to generate candidates, evaluate them, and learn from the results. The quality of this feedback is critical: without reliable external evaluation, LLMs may reinforce their own reasoning errors~\cite{huang2024large}, while agents optimized against tests may learn to exploit the evaluator~\cite{sakana2025dgm}. Our framework grounds this feedback in formal verification. Each accepted refinement step is checked against its proof obligations, and unsuccessful attempts provide diagnostic information to guide further exploration. The resulting derivation traces could support model training, prompt optimization, and the improvement of refinement search strategies, while verified components could accumulate in a reusable library~\cite{wang2023voyager}.
These improvements would enable the system to generate further verified experience and use it to improve the mechanisms that guide subsequent derivations, forming a path toward recursive self-improvement.

The main contributions of this paper are as follows.

\noindent $\bullet$ A contract language for synchronous hardware embedded in Dafny, together with an auto-formalization loop that translates natural-language specifications into machine checked and semantically validated contracts.

\noindent $\bullet$ A refinement calculus for synchronous hardware comprising 26 core laws and 11 compositional laws derived from them.
Each law combines a design decision, its application conditions, and a mechanical transformation.

\noindent $\bullet$ A comprehensive toolchain that consults the LLM only for design decisions, verifies each refinement step with Dafny, and translates the verified implementation into RTL with line-by-line traceability.

\noindent $\bullet$ An evaluation on VerilogEval, RTLLM, and ChipBench~\cite{liu2023verilogeval,lu2024rtllm,yu2026chipbench} showing that our framework achieves pass rates comparable to agentic baselines.
More importantly, by conducting formal verification at each design step, our framework ensures that every generated design satisfies its contract, and an LLM agent cannot inflate the pass rate by overfitting to defective testbenches.

%% file: src/background.tex
\section{Background and Problem Formulation}
\label{sec:background}

\noindent \textbf{LLM-Based RTL Generation.}
Agentic systems accomplish complex tasks by coordinating planning, reasoning, and tool use. RTL generation systems typically operate in a closed feedback loop: agents generate a design, compile and simulate it, and iteratively revise the code based on diagnostic feedback~\cite{yang2026large,tsai2024rtlfixer,zhao2025mage}.
However, a failed test reveals only that an error exists, not where it originates. Therefore, recent systems allow agents to inspect waveforms and trace failures back to the signals and program locations that caused them~\cite{ho2025verilogcoder,liu2026veritrace}. Post-training further internalizes this feedback loop by rewarding models for generating designs that pass a testbench~\cite{zhu2025codevr1,teng2025verirl}. Nevertheless, the testbench remains the ultimate correctness oracle, and its coverage is inherently incomplete: 44.2\% of designs that pass the testbenches of a widely used benchmark still fail formal verification~\cite{jin2025realbench}. Moreover, some benchmark cases are themselves flawed~\cite{wang2026rtlbenchmt}, and strong performance on small modules does not necessarily transfer to realistic designs~\cite{yu2026chipbench}.

There is an increasing trend to incorporate structural information into the generation process, as generating an entire design in one pass becomes more brittle as the design complexity grows. 
One family frames generation as a search problem: it samples candidate designs, performs agentic workflows, and retains the most promising candidates~\cite{chao2026veribtot,min2026revolution,wei2026vflow,ramamoorthy2026verimaas,vijayaraghavan2026stepprm}. 
Another family introduces an intermediate representation before RTL, such as a structured behavioral description, a statement of design intent, or a decomposition into submodules, and then lowers it to code~\cite{huang2026crux,kong2026verirefine,cheng2026hint,lyu2026chipv}.
Both approaches make the generation process more explicit, but neither makes it formally verifiable. Search-based methods still rely on incomplete testbenches to rank candidates, while intermediate representations above lack formal semantics and are still judged by the same model that produced them.

Formal methods have been integrated into LLM-based RTL generation in two ways.
As checkers, they prove equivalence against a software reference model, verify assertions on completed RTL, or provide reward signals for training~\cite{li2026formalrtl,liu2026verisure,shi2026earl}.
As generation languages, they allow models to produce typed hardware descriptions in a proof assistant or verified software that can later be translated into hardware~\cite{xiong2026cktformalizer,jha2025proof2silicon}.
However, neither approach makes the design process correct by construction.
Post-hoc verification requires a golden reference model, which is often unavailable for a new design. Type checking can rule out structural errors, but it does not guarantee functional correctness.
No prior work makes each intermediate design decision formally verifiable, thereby establishing correctness throughout RTL construction rather than only after generation.

\noindent \textbf{High-Level Synthesis.}
High-level synthesis (HLS) can automatically translate a behavioral model, typically written in C or C++, into an RTL implementation~\cite{cong2011high}. Because HLS requires a complete and executable behavioral model, it cannot directly synthesize RTL from an incomplete specification that states only design requirements. To bridge this gap, recent work uses LLMs to transform existing software into HLS-compatible code~\cite{collini2025c2hlsc,xiong2024hlspilot} or to translate specification documents into C++ models~\cite{yu2025spec2rtl}. However, introducing such a behavioral model adds an extra layer of abstraction, which may deviate from the original specification and obscure hardware-specific design decisions.

\noindent \textbf{Program Refinement.}
Program refinement derives an executable program from an abstract specification through a sequence of verifiable steps~\cite{wirth1971program,dahl1972structured}. A \textit{refinement calculus} provides a collection of sound laws for these steps, so that a program derived entirely through such laws is correct by construction~\cite{back1988calculus,morris1987theoretical,morgan1990programming}.
Researchers have developed refinement calculi for a range of domains, including concurrent systems~\cite{breuer1997refinement,dingel2002refinement}, reactive systems~\cite{preoteasa2014refinement,hallerstede2003specification}, real-time systems~\cite{hayes2001sequential}, and state charts~\cite{scholz1998refinement,rumpe1996automata}.
Although tools can partially automate program transformation and mechanical proof~\cite{bauer1989formal,lammich2019refinement,leino2010dafny}, refinement still relies on human expertise to select suitable laws and construct artifacts such as invariants, guards, and expressions.
LLMs provide a natural way to automate these creative choices. Recent approaches use LLMs to select refinement laws and generate the corresponding artifacts~\cite{cai2025automated,wang2026eventb}. However, these software-oriented approaches do not support hardware semantics, such as concurrent processes, simultaneous register updates, and cycle-by-cycle signal behavior.

\noindent \textbf{Problem Definition.}
Given a design specification expressed in natural language, tables, or figures, we aim to generate a synthesizable RTL implementation through interpretable and verifiable steps.
The formal contract should faithfully capture the requirements, and the implementation must be proved to satisfy the contract.

%% file: src/specification.tex
\section{Formal Specification of Hardware}
\label{sec:specification}

\begin{figure*}[tp]
\centering
\footnotesize
\setlength{\grammarparsep}{3pt}
\setlength{\grammarindent}{4.4em}
\begin{minipage}[t]{0.53\textwidth}
\begin{grammar}
<term> ::= <id> | <id>$'$ | <const> | <op>\texttt{(}<term-list>\texttt{)} | <id>\texttt{(}<term-list>\texttt{)}

<pred> ::= \texttt{true} | \texttt{false} | <relational-op>\texttt{(}<term-list>\texttt{)}

<form> ::= <pred> | <logical-op>\texttt{(}<form-list>\texttt{)}
\alt \texttt{forall} <id> $\bullet$ <form> | \texttt{exists} <id> $\bullet$ <form>
\alt \texttt{X} <form> | \texttt{Y} <form> | \texttt{G}$[l,u]$ <form> | \texttt{H}$[l,u]$ <form>
\alt \texttt{F}$[0,u]$ <form> | <form> \texttt{U}$[0,u]$ <form>

<spec> ::= <id-list>\texttt{:[pre:}<form>\texttt{, dur:}<form>\texttt{, post:}<form>\texttt{]} \texttt{\char124} \texttt{env:}<form>

<sig> ::= \texttt{module} <id>\texttt{(}<id-list>\texttt{)} \texttt{returns} \texttt{(}<id-list>\texttt{)}

<module> ::= <sig> <spec>
\end{grammar}
\end{minipage}%
\hfill
\begin{minipage}[t]{0.45\textwidth}
\begin{grammar}
<expr> ::= <id> | <const> | <op>\texttt{(}<expr-list>\texttt{)} | <id>\texttt{(}<expr-list>\texttt{)}
\alt \texttt{if} <expr> \texttt{then} <expr> \texttt{else} <expr>

<func> ::= \texttt{function} <id>\texttt{(}<id-list>\texttt{)} \texttt{\{}<expr>\texttt{\}}

<stmt> ::= \texttt{skip} | <stmt>\texttt{;} <stmt>
\alt \texttt{let} <id> \texttt{:=} <expr> | <id> \texttt{:=} <expr> | <id>$'$ \texttt{:=} <expr>
\alt \texttt{if} <expr> \texttt{then} <stmt> \texttt{else} <stmt>
\alt <id-list> \texttt{:=} <id>\texttt{(}<expr-list>\texttt{)}

<sig> ::= \texttt{module} <id>\texttt{(}<id-list>\texttt{)} \texttt{returns} \texttt{(}<id-list>\texttt{)}

<module> ::= <sig> <spec> <stmt>
\alt <func> <module> | <module> <module>
\end{grammar}
\end{minipage}
\caption{Core syntax of the specification language \Lspec{} (left) and the implementation language \Lcode{} (right).}
\label{fig:syntax}
\end{figure*}

In this section, we introduce two complementary formal languages:
\Lspec{} for specifying hardware contracts and \Lcode{} for
describing their implementations.
To facilitate LLM-based generation and reasoning, we define both
languages as subsets of Dafny~\cite{leino2010dafny}, a widely used
programming language for formal verification.
Together, they enable stepwise refinement from abstract contracts
to concrete implementations through intermediate versions
that combine specifications and code, as detailed in
Section~\ref{sec:refinement}.

\noindent\textbf{Specification Language.}
A \textit{specification statement} represents an abstract program~\cite{morgan1990programming}. During refinement, it denotes a program fragment that remains to be developed. In the software domain, it takes the form $w:[\mathit{pre}, \mathit{post}]$, where the precondition constrains the initial state, the postcondition constrains the final state, and the frame $w$ lists the variables that may change their values. From any initial state satisfying the precondition, the fragment must terminate in a state satisfying the postcondition, while modifying only the frame variables. However, a hardware module may run indefinitely, so its behavior must be specified throughout execution~\cite{breuer1997refinement}. In addition, its correct operation may depend on assumptions about the environment. We therefore extend the statement with a \textit{during condition} and an \textit{environment condition}, yielding the form $w:[\mathit{pre}, \mathit{dur}, \mathit{post}] \mid \mathit{env}$ (Fig.~\ref{fig:syntax}). The program fragment must maintain $\mathit{dur}$ throughout execution and may rely on $\mathit{env}$ as an assumption during refinement~\cite{jones1983tentative,abadi1993composing}. When refining the body of a single process, we incorporate the environment condition into the during condition, obtaining the simplified form $w:[pre,dur,post]$. At this level, the combined condition provides the context for local development.

\noindent\textsc{Example.}
Fig.~\ref{fig:gcd} presents our running example.
The contract specifies \textit{what} the module computes without prescribing \textit{how}.
The function \texttt{gcd} is defined declaratively, while the during condition requires the greatest common divisor of the two registers to remain unchanged.
The choice of algorithm, such as Euclid's repeated subtraction, is left to refinement.
The last sentence of the design specification follows directly from the during condition: once the registers become equal, their common value must be $\mathtt{gcd}(\mathit{init}(a),\mathit{init}(b))$.
The postcondition is \texttt{false} because the module never terminates, and the environment condition is \texttt{true} because the module is standalone.

To keep the example concise, we make two simplifications.
First, the registers also serve as outputs and hold the operands at power-up, eliminating the need for a loading protocol.
Second, the contract omits the progress requirement that the registers eventually become equal.
Instead, we establish progress separately using a ranking function that strictly decreases until the registers become equal.

\begin{figure}[tp]
\begin{lstlisting}[style=promptstyle]
 - output a (8 bits), b (8 bits)

The module computes the greatest common divisor of two positive numbers initially stored in registers a and b. Throughout the computation, the greatest common divisor of a and b must remain equal to that of their initial values. When the two registers become equal, their common value is the result.
\end{lstlisting}
\begin{lstlisting}[style=specstyle, columns=flexible, basewidth=0.52em]
gcd(x, y) == max {d : x mod d = 0 $\wedge$ y mod d = 0}

euclid == a, b : [pre:  a > 0 $\wedge$ b > 0,
                  dur:  gcd(a, b) = gcd(init(a), init(b)),
                  post: false] | env: true
\end{lstlisting}
\caption{The design specification (top) and the \Lspec{} contract (bottom) of the \texttt{euclid} example.}
\label{fig:gcd}
\end{figure}

\noindent\textbf{Implementation Language.}
We define a simple language \Lcode{} to represent hardware implementations.
A logic circuit consists of interconnected \textit{modules}, each evaluated once per clock cycle.
Within each cycle, the \textit{statements} of a module are evaluated sequentially. A let binding introduces an intermediate value that synthesizes to a combinational wire, while an output assignment drives an output port and takes effect immediately. A register update, written \texttt{<id>' := <expr>}, schedules a new value for the target register based on the current value of the expression. At the next clock edge, all scheduled register updates take effect simultaneously, while registers without scheduled updates retain their values. Modules may instantiate verified submodules, and \textit{functions} provide reusable combinational logic.
Explicit delays are not permitted because they are not synthesizable.
The language is functionally complete, as it can represent any concrete behavior specified by the formal specification language~\cite{breuer1997refinement}.
A rule-based generator translates verified \Lcode{} programs into SystemVerilog, Verilog, and VHDL, preserving traceability between each line of generated RTL and its originating \Lcode{} statement.

%% file: src/refinement.tex
\section{LLM-Guided Hardware Refinement}
\label{sec:refinement}

Stepwise refinement develops a program from its specification through a sequence of small, verifiable steps~\cite{wirth1971program,back1988calculus}:

\begin{equation}
\mathit{spec} \sqsubseteq \mathit{ver}_1 \sqsubseteq \cdots \sqsubseteq \mathit{ver}_n \sqsubseteq \mathit{code},
\label{eqn:refinement}
\end{equation}

where each intermediate version may combine specification statements with concrete code. A version $P$ is refined by $Q$, written $P \sqsubseteq Q$, if every behavior permitted by $Q$ is also permitted by $P$. In other words, the refinement process progressively reduces nondeterminism while preserving all original requirements~\cite{morris1987theoretical,morgan1988specification}.

In practice, the refinement process is guided by a set of predefined \textit{refinement laws}.
In our framework, they serve as the interface between the LLM and the refinement engine. Each law consists of three components: a \textit{design decision}, a set of \textit{application conditions}, and a \textit{transformation}. The design decision specifies a choice to be made, such as selecting a predicate, constructing an invariant, or determining a value, and provides a template for the LLM to fill in. The application condition defines the proof obligations that the proposed choice must satisfy, which are checked by Dafny before the law can be applied. Once these obligations are discharged, the refinement engine performs the corresponding transformation mechanically.
This separation allows the LLM to make creative design decisions without directly modifying the program.
Since only proposals that satisfy the proof obligations are accepted, correctness rests on the soundness of the refinement laws rather than the reliability of the LLM.
Moreover, recording each refinement step allows the user to trace every line of the implementation back to the design decision that produced it.

\noindent\textsc{Example.}
Consider the \textit{Iteration} law, which reduces the task of refining an indefinitely running process to refining its behavior over a single clock cycle. 
As shown in box 1 of Fig.~\ref{fig:euclid}, the LLM makes two design decisions when applying this law: implementing \texttt{euclid} as a single process rather than multiple concurrent processes, and choosing an inductive invariant $\mathit{inv}$ for that process. Here, the LLM proposes the conjunction of $\mathit{pre}$ and $\mathit{dur}$ in Fig.~\ref{fig:gcd} as $\mathit{inv}$. The application conditions require that $\mathit{inv}$ hold in the reset state and imply $\mathit{dur}$, both of which can be verified by Dafny. The transformation produces a specification statement for the process body, which assumes $\mathit{inv}$ at the start of each clock cycle and requires it to hold again at the end.

\noindent\textbf{A Refinement Calculus for Hardware Design.}
Despite the extensive literature on refinement calculus (Section~\ref{sec:background}), existing approaches cannot be directly applied to LLM-guided hardware design for two reasons. First, they primarily target terminating sequential programs, whereas
hardware modules may run indefinitely and must satisfy their requirements
throughout execution. Second, they rely on human experts to select and instantiate refinement laws, whose formal semantics are not readily accessible to LLMs. To address these challenges, we develop a hardware-oriented refinement calculus comprising 26 core laws. 
Due to space limits, we do not enumerate them here.
Instead, we describe a common design pattern that progressively transforms a contract into a concrete implementation:

\noindent $i)$ \textit{Temporal Unfolding.}\
The first stage dissolves temporal operators from the contract, so that every later stage reasons about an equivalent step relation. Following the expansion laws of temporal logic~\cite{manna1990hierarchy}, each operator is unfolded into a condition over the current and the next states, with auxiliary variables introduced as needed~\cite{abadi1991existence}. The translation is fully automatic and involves no design decisions.

\noindent $ii)$ \textit{Structural Decomposition.}
A complex module can be divided into smaller ones that are refined independently. The LLM decides how to partition the module into concurrent processes or
delegate parts of its functionality to known implementations. The $\mathit{dur}$ of one process serves as $\mathit{env}$ for others,
thus enables compositional verification~\cite{abadi1993composing}.

\noindent $iii)$ \textit{Loop Extraction.}
As discussed in the preceding example, this stage reduces the refinement of a non-terminating process to that of its one-cycle body. The LLM selects an inductive invariant, and the resulting specification statement assumes it at the beginning of each cycle and requires it to be re-established at the end. By induction over clock cycles, verifying the process body is sufficient to establish the correctness throughout execution.

\noindent $iv)$ \textit{Mechanical Realization.}
This stage applies refinement laws automatically, without requiring LLM decisions. Normalization laws simplify and restructure specification statements while preserving their semantics, whereas realization laws convert behavior already determined by the contract into concrete code. The refinement engine repeatedly applies these laws until no further transformations are possible. Any remaining underdetermined behavior is left as a \textit{typed hole}, a specification statement that identifies what remains to be implemented and what design decision is needed.

\noindent $v)$ \textit{Guided Decision.}
This stage uses the LLM to resolve design choices left open by specification statements. For each typed hole, the refinement engine presents a specific question to the LLM, such as how to structure control flow, strengthen logical conditions, or introduce intermediate variables and state. Rather than selecting a refinement law directly, the LLM provides an answer that the engine maps to an appropriate law. Dafny verifies the corresponding application conditions before the transformation is performed. \textit{Mechanical Realization} and \textit{Guided Decision} alternate until all specification statements have been replaced by concrete code.

\begin{figure}[tp]
\begin{lstlisting}[style=codestyle]
euclid
$\sqsubseteq$ "Iteration" with inv == a > 0 $\wedge$ b > 0 $\wedge$
        gcd(a, b) = gcd(init(a), init(b))
module euclid() returns (a, b) {
  a, b : [pre: inv, dur: true, post: inv'](*@\hfill\texttt{<--}@*)
}
\end{lstlisting}
\begin{lstlisting}[style=codestyle]
$\sqsubseteq$ "Alternation" with guards a > b, a < b, a = b
if a > b then
  a, b : [pre: inv $\wedge$ a > b, dur: true, post: inv'](*@\hfill\texttt{<--}@*)
else if a < b then
  a, b : [pre: inv $\wedge$ a < b, dur: true, post: inv']
else
  a, b : [pre: inv $\wedge$ a = b, dur: true, post: inv']
\end{lstlisting}
\begin{lstlisting}[style=codestyle]
$\sqsubseteq$ "Strengthen Postcondition" with ranking a + b
a, b : [pre: inv $\wedge$ a > b, dur: true,
        post: inv' $\wedge$ a' + b' < a + b](*@\hfill\texttt{<--}@*)
$\sqsubseteq$ "Assignment"
a' := a - b
\end{lstlisting}
\begin{lstlisting}[style=specstyle]
module euclid() returns (a, b) {
  if a > b then a' := a - b
  else if a < b then b' := b - a
  else skip
}
\end{lstlisting}
\caption{Stepwise refinement of the \texttt{euclid} example.
From top to bottom, the \textit{Iteration} law extracts the
one-cycle body from the process in Fig.~\ref{fig:gcd}, the \textit{Alternation} law splits it into
three branches, and the first branch is refined to ensure
that the ranking function $a+b$ strictly decreases.
The second branch is refined similarly, while the third is
completed with a \texttt{skip} statement.
The arrow indicates the statement to be refined next.}
\label{fig:euclid}
\end{figure}

\noindent\textsc{Example.}
Fig.~\ref{fig:euclid} illustrates how the \texttt{euclid} contract (Fig.~\ref{fig:gcd}) is refined into a concrete implementation.
The first two stages leave the contract unchanged because it contains no temporal operators and the LLM chooses to implement it as a single process rather than two concurrent processes.
In stage \textit{iii)}, the \textit{Iteration} law extracts the one-cycle process body using the inductive invariant chosen in the preceding example (box 1).
In stage \textit{iv)}, mechanical realization cannot proceed because the invariant does not determine the next value of either register, leaving the entire body as a typed hole.
In stage \textit{v)}, the LLM proposes three guards based on the relative values of the registers.
After Dafny verifies that the guards cover all possible cases, the \textit{Alternation} law splits the body into three branches, adding each guard to the corresponding branch's precondition (box 2).
To ensure progress, the LLM proposes a ranking function $a+b$ that strictly decreases until the registers become equal.
The \textit{Strengthen Postcondition} law adds this requirement to the postcondition of the first branch.
The LLM then proposes $a-b$ as the next value of register $a$, which mechanical realization converts into the assignment $a' := a-b$ (box 3).
Dafny verifies that this update preserves the invariant and satisfies the strengthened postcondition.
The second branch is refined similarly, while the third is completed with a \texttt{skip} statement because the registers already hold the result.
The final implementation (box 4) contains no remaining specification statements.

Our refinement laws are derived from established principles of formal program refinement.
Laws operating within a process body follow the weakest-precondition semantics of classical refinement calculi~\cite{dijkstra1975guarded,morgan1990programming}.
For hardware-specific constructs, we build on mathematical induction over repeated process execution~\cite{breuer1997refinement}, assume-guarantee reasoning for concurrent process composition~\cite{jones1983tentative,abadi1993composing}, and data refinement for introducing or modifying state representations~\cite{hoare1972proof}.
Each law is derived from these principles, with application conditions specifying when its transformation preserves correctness.

\noindent\textbf{Hardware Design as a Search Problem.}
Given an initial contract and a refinement calculus, the possible derivations of a hardware design form a \textit{refinement tree}.
The root represents the original contract, each edge corresponds to a refinement law application, and each node represents a program version.
The leaves are either completed implementations or infeasible versions that cannot be refined further.
Hardware design thus becomes a search problem of finding a path from the root to a completed implementation.
Two features of our framework make this search tractable for LLMs.
First, deterministic normalization reduces the search space by automatically performing transformations that require no design decisions, allowing the LLM to focus on questions associated with typed holes.
Second, formal verification provides dense and reliable feedback throughout the refinement process.
Invalid proposals are rejected with diagnostic information, allowing the LLM to revise its decisions or backtrack to explore alternative paths.
Moreover, the refinement trace records each design decision and its formal justification, making the development process interpretable and traceable.
Since functional correctness is enforced independently of the search strategy, future work could further optimize the search for power, performance, and area without compromising correctness.

\noindent\textsc{Example.}
Consider the following scenario: after extracting the process body in Fig.~\ref{fig:euclid}, the LLM attempts to ensure progress by strengthening the postcondition of the entire body with $a'+b'<a+b$.
However, when $a=b$, no register update can strictly decrease $a+b$ while preserving the inductive invariant.
Dafny identifies this conflict through a counterexample, prompting the framework to backtrack to the original process body.
Guided by this feedback, the LLM first splits the body into three branches and requires the ranking function to decrease only when the registers differ, as illustrated in Fig.~\ref{fig:euclid}.

%% file: src/implementation.tex
\begin{figure*}[!t]
    \centering
    \centerline{\includegraphics[width=1.00\linewidth]{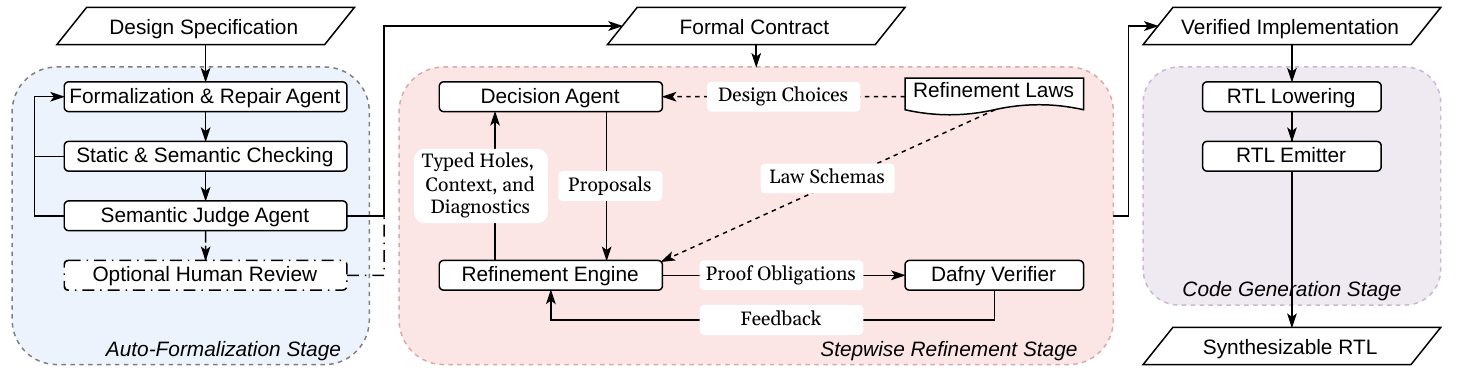}}
    \caption{Overview of the hardware generation workflow.}
    \label{fig:system}
    \vspace{-3pt}
\end{figure*}

\section{Correct-by-Construction RTL Generation}
\label{sec:system}

We implement the techniques described in the preceding sections as an end-to-end workflow comprising three stages (Fig.~\ref{fig:system}).
The auto-formalization stage translates design specifications into formal contracts, the stepwise refinement stage derives verified programs from these contracts, and the code generation stage translates the programs into RTL.
Beyond producing a verified implementation, the workflow maintains a complete record of the design process: each contract clause is linked to its original requirement, each refinement step records the design decision and its formal justification, and each line of RTL is traced to the statement and refinement law that produced it.
This traceability allows engineers to understand how the implementation was derived and pinpoint potential errors.

\noindent\textbf{Auto-Formalization Stage.}
The first stage translates an informal design specification into a formal contract in \Lspec{}.
Because natural language can be ambiguous or incomplete, formal verification alone cannot establish whether the contract faithfully captures the intended requirements.
We therefore combine automated validation, iterative repair, and human review to improve the reliability of this translation.
The generated contract undergoes three kinds of validation (Fig.~\ref{fig:system}), each targeting a different class of errors.
\textit{Static checks} detect syntax errors and constructs unsupported by hardware synthesis.
\textit{Semantic probes} use Dafny to verify properties of the contract itself, including the consistency of its clauses, whether every output and register is constrained, and agreement with formally checkable information from tables and diagrams in the specification.
Finally, an LLM-based \textit{semantic judge} compares a back-translation of the contract with the original specification, using mechanically computed facts to support its assessment.
If any check fails, the LLM revises the contract based on diagnostic feedback, and the validation process repeats within a bounded retry budget.
Since automated validation cannot fully resolve ambiguities in the original specification, an engineer may review and amend the contract before refinement begins.
The contract is designed for easy review: it is concise, its clauses use syntax resembling SystemVerilog Assertions, and each clause is labeled with a reference to the requirement it formalizes.

\noindent\textbf{Stepwise Refinement Stage.}
The second stage derives a verified implementation from the formal contract through the refinement search described in Section~\ref{sec:refinement}. To minimize the burden on the LLM, the refinement engine handles all mechanical transformations, while the LLM-based decision agent is consulted only when a design decision is needed (Fig.~\ref{fig:system}). For each typed hole, the engine presents a focused question, together with the relevant constraints, target types, and available variables.
The decision agent responds by proposing a value, a case split, an invariant, an intermediate condition, a submodule with its contract, etc. The engine then selects the corresponding refinement law, generates its proof obligations for Dafny, and applies the transformation only after those obligations are discharged. In this way, the decision agent makes design decisions without directly selecting refinement laws or modifying the program.

To explore alternative designs, the decision agent is requested to generate several ranked proposals for each typed hole.
Each proposal is checked before acceptance, and rejected proposals return diagnostic feedback to guide subsequent attempts.
The search proceeds along the most promising refinement path while retaining alternatives for backtracking if a dead end is encountered.

\noindent\textbf{Code Generation Stage.}
The third stage translates the verified program into RTL.
Because \Lcode{} is designed to match the clocked semantics of RTL, this translation is syntax-directed and requires no further design decisions.
The compiler first lowers the program into a target-independent hardware representation, discarding annotations used only for verification.
Language-specific emitters then generate SystemVerilog, Verilog, or VHDL.
To preserve traceability, the compiler also produces a correspondence map linking each line of RTL to the program statement and refinement law that produced it.
Together with the contract's links to the original requirements and the recorded refinement steps, this allows engineers to trace each line of generated RTL back to the requirement it implements.

%% file: src/evaluation.tex
\section{Evaluation}
\label{sec:evaluation}

\begin{table*}[t]
\centering
\caption{Number of problems that pass the golden testbench and total token consumption for each method. Each cell gives the values with Gemini 3.8 Flash and Gemini 4 Argon, separated by a slash.}
\label{tab:main}
\setlength{\tabcolsep}{4.8pt}
\begin{tabular}{@{}llccccccccc@{}}
\toprule
\multirow{3}{*}[-0.5ex]{Testbench} & \multirow{3}{*}[-0.5ex]{Method} & \multicolumn{4}{c}{Defect-free} & \multicolumn{4}{c}{Defective} & \multirow{3}{*}[-0.5ex]{\begin{tabular}[c]{@{}c@{}}Total\\tokens\\(M)\end{tabular}} \\
\cmidrule(lr){3-6}\cmidrule(lr){7-10}
 & & \textit{VerilogEval} & \textit{RTLLM} & \textit{ChipBench} & Total & Contradict. & Unreliable & Ambiguous & Total & \\
 & & (152) & (42) & (16) & (210) & (9) & (9) & (8) & (26) & \\
\midrule
\multirow{3}{*}{\textit{w/o access}} & Plain LLM & 145/147 & 39/39 & 14/14 & 198/200 & 0/1 & 4/5 & 2/5 & 6/11 & 0.8/0.2 \\
 & MAGE (hidden) & 150/150 & 39/37 & 14/15 & 203/202 & 0/1 & 6/5 & 1/5 & 7/11 & 7.9/4.4 \\
 & Ours (end-to-end) & 149/149 & 39/38 & 14/14 & 202/201 & 0/0 & 1/3 & 1/4 & 2/7 & 4.8/2.6 \\
\midrule
\multirow{3}{*}{\textit{w/ access}} & MAGE & 152/152 & 42/42 & 15/16 & 209/210 & 6/8 & 7/8 & 6/7 & 19/23 & 14.6/5.7 \\
 & ACE-RTL & 152/152 & 41/42 & 16/16 & 209/210 & 4/8 & 5/6 & 6/7 & 15/21 & 6.1/1.5 \\
 & Ours (amended) & 150/150 & 41/40 & 14/14 & 205/204 & 6/6 & 3/5 & 7/7 & 16/18 & 4.6/2.6 \\
\bottomrule
\end{tabular}
\vspace{-3pt}
\end{table*}

\noindent\textbf{Experimental Setup.}
We evaluate our framework on three benchmark suites: VerilogEval~\cite{liu2023verilogeval}, RTLLM v2.1~\cite{lu2024rtllm}, and the self-contained tier of ChipBench~\cite{yu2026chipbench}.
We use two base models, Gemini 3.8 Flash and Gemini 4 Argon, both with their default reasoning effort.
We compare against three baselines: (1) a plain baseline that generates Verilog in a single LLM call; (2) MAGE~\cite{zhao2025mage}, a multi-agent system that generates its own testbench and uses it to debug the design; and (3) ACE-RTL~\cite{deng2026ace}, which iterates among a generator, a reflector, and a coordinator.
The plain baseline is instructed to generate Verilog-2001, while the agentic baselines are run with their default configurations, except for an increased output token limit.
All experiments are conducted on a MacBook equipped with an M5 Pro processor and 24\,GB of memory.
We use Dafny 4.11.0 and Z3 5.3.3 for verification, and Icarus Verilog 13.0 for simulation.

\noindent\textbf{Testbench Overfitting on Defective Benchmarks.}
Each problem in the three benchmark suites includes a reference design and a golden testbench.
For VerilogEval and ChipBench, the golden testbench determines the expected outputs by simulating the reference design, whereas RTLLM encodes the expected behavior directly in the testbench.
By default, both MAGE and ACE-RTL use feedback from the golden testbench to iteratively improve their generated designs.
Consequently, their final designs are evaluated against the same testbench used for generation, creating a risk of overfitting to the testbench rather than satisfying the intended design requirements.

However, the benchmarks themselves contain defects.
We audited all three suites and identified 26 problems in which benchmark defects affect the evaluation verdict.
We classify these problems into three categories based on their primary defects.
A problem is \emph{contradictory} (\textit{VerilogEval 62, 93, 149; ChipBench 4--7, 15, 33}) if the prompt specifies one behavior but the testbench expects another, so a design can pass only by deviating from the prompt.
A problem is \emph{unreliable} (\textit{RTLLM} \texttt{clk\allowbreak{}generator}, \texttt{fixed\_\allowbreak{}point\_\allowbreak{}adder}, \texttt{fixed\_\allowbreak{}point\_\allowbreak{}substractor}, \texttt{freq\_\allowbreak{}divbyeven}, \texttt{radix2\_\allowbreak{}div}; \textit{ChipBench 8--10, 13}) if the testbench is faulty and produces verdicts that do not reliably reflect the correctness of the design.
For example, the testbench may reject even the reference design or introduce a race condition that makes the verdict depend on the simulator's event ordering.
A problem is \emph{ambiguous} (\textit{VerilogEval 153; RTLLM} \texttt{pulse\_\allowbreak{}detect},\texttt{ring\_\allowbreak{}counter},\texttt{serial2\allowbreak{}parallel}; \textit{ChipBench 2, 17, 27, 34}) if the prompt leaves some behavior unspecified or describes it inconsistently, while the testbench accepts only one possible interpretation.

\noindent\textbf{Main Results.}
Table~\ref{tab:main} reports the number of problems that pass the golden testbench.
The 26 defective problems are counted separately from the 210 defect-free ones.
The methods are grouped by whether they have access to the golden testbench during design generation.

We first compare the methods that do not have access to the golden testbench during design generation.
The three methods achieve similar pass rates.
The strong performance of plain LLM can be partly attributed to data contamination~\cite{wang2025vericontaminated}.
In contrast, our method cannot simply reproduce a memorized reference design, as stepwise refinement is a new task for LLMs.
Moreover, our method can provide a formal proof that each generated design satisfies its contract.

The strength of our method is most apparent on defective problems. In a separate experiment, we explicitly prompt the LLM to enumerate the open behaviors in the contract and to generate diverse designs across five attempts.
On contradictory problems, none of the five designs passes the golden testbench, since our method guarantees that every generated design satisfies its formal contract.
On ambiguous problems, at least two of the five designs exhibit non-equivalent behaviors, revealing ambiguity in the specification.
Unreliable problems are harder to detect because the defects lie in the testbench rather than in the specification.
Plain LLM and MAGE with hidden testbenches pass more defective problems (Table~\ref{tab:main}), but their generated designs are not guaranteed to satisfy the intended requirements.

Next, we compare our method with the baselines that have access to the golden testbench. We amend the prompt of each problem with the behavior required by the reference design and the golden testbench, and let the testbench override the original requirements when there is a contradiction.
Under this setting, our method also overfits to the testbench, achieving pass rates comparable to those of the baselines.
Across all problems and settings, Dafny solving accounts for only 4\% and 13\% of the total LLM generation and solver time with Flash and Argon, respectively.

\noindent\textbf{Limitations.}
First, despite advances in auto-formalization, our framework still relies on human engineers to validate that the generated contracts accurately capture the intended design requirements.
Second, multi-clock designs must be decomposed into single-clock modules.
Additionally, our current framework does not support level-sensitive storage, which accounts for two failures in our evaluation.

\begin{figure}[t]
\centering
\includegraphics[width=0.8\linewidth]{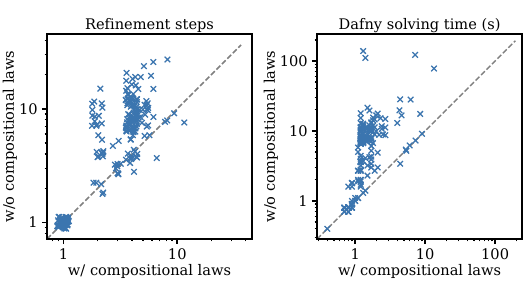}
\caption{Comparison of refinement steps (left) and Dafny solving time (right) between core laws only and core laws with compositional laws.}
\label{fig:ablation}
\vspace{-3pt}
\end{figure}

\noindent\textbf{Toward Self-Improvement with Compositional Laws.}
A compositional law packages a recurring sequence of core-law steps into a single law.
Since it is derived from core laws, it maintains the proving power of the calculus while recording design experience in a reusable form~\cite{cai2025automated}.
To measure its benefit, we refine single-clock problems with and without the 11 compositional laws.
With Gemini 4 Argon, both settings succeeded on 224 problems.
However, as Fig.~\ref{fig:ablation} shows, without the compositional laws the refinement takes 2.1$\times$ as many steps and 5.1$\times$ as much Dafny solving time in total.
Compositional laws thus make later derivations more efficient without enlarging the trusted core.
Extracting them from verified derivations is a promising path toward self-improvement.

%% file: src/conclusion.tex
\section{Conclusion}

In this paper, we proposed a comprehensive framework for LLM-guided, correct-by-construction hardware generation.
Beyond generating verified implementations, this framework establishes a foundation for LLM-native hardware design, enabling systematic design exploration with reliable verification feedback and providing opportunities to accumulate reusable design knowledge.
Future directions include incorporating power, performance, and area objectives into the refinement search and leveraging derivation traces to improve LLMs and search strategies.